\documentclass{pasj00}

\newcommand{\fx}{\ensuremath{F_{\rm{X}}}}
\newcommand{\nh}{\ensuremath{N_{\rm H}}}

\newcommand{\axj}{AX~J1745.6--2901}
\newcommand{\gs}{GS~1741.2--2859/1741.6--2849}
\SetRunningHead{Y.~Hyodo et al.}{Suzaku Study of 
AX J1745.6--2901}
\Received{2008/7/12}
\Accepted{2008/8/31}
\begin{document}
\title{Timing and Spectral Study of AX\,J1745.6--2901 with Suzaku}
\author{Yoshiaki \textsc{Hyodo},\altaffilmark{1} 

Yoshihiro \textsc{Ueda},\altaffilmark{2} 
Takayuki \textsc{Yuasa},\altaffilmark{3}\\ 
Yoshitomo \textsc{Maeda},\altaffilmark{4}
Kazuo \textsc{Makishima},\altaffilmark{3}
and 
Katsuji \textsc{Koyama},\altaffilmark{1} 
}
\altaffiltext{1}{Department of Physics, Graduate School of Science, 
Kyoto University, \\Kita-shirakawa Oiwake-cho, Sakyo, Kyoto 606-8502}
\altaffiltext{2}{Department of Astronomy, Graduate School of Science, 
Kyoto University, \\Kita-shirakawa Oiwake-cho, Sakyo, Kyoto 606-8502}
\altaffiltext{3}{Department of Physics, School of Science, 
the University of Tokyo, 7-3-1, Hongo, Bunkyo, Tokyo 113-0033}
\altaffiltext{4}{Institute of Space and Astronautical Science, 3-1-1, 
Yoshinodai, Sagamihara, Kanagawa, 229-8510}
\email{hyodo@cr.scphys.kyoto-u.ac.jp}
\maketitle 
\begin{abstract}
The eclipsing low-mass X-ray binary \axj\ was observed with Suzaku in its
outburst phase. 
Combining the Chandra observation made 1.5 month earlier than Suzaku, 
the orbital period is determined to be  $30063.76\pm$0.14~s.
We found deep flux dips prior to the eclipse phase of orbit. 
The X-ray spectrum in the persistent phase is described with 
a combination of a direct and a scattered-in by dust emissions. 
During the eclipse, the X-ray spectrum becomes only the dust scattering 
(scattered-in) component. 
The direct component is composed of a disk-blackbody and 
a blackbody (neutron star surface). 
No power-law component is found in the hard energy band up to 30~keV. 
The optical depth of the dust-scattering is $\sim10.5$ at 1~keV. 
A clear edge at $\sim 7.1$~keV in the deep dip spectrum indicates that the 
major portion of Fe in the absorber is neutral or at low ionization state. 
We discovered four narrow absorption lines near the K-shell transition 
energies of Fe\emissiontype{XXV}, Fe\emissiontype{XXVI}, and 
Ni\emissiontype{XXVII}. 
The absorption line features are well explained by the solar abundance gas in 
a bulk motion of $\sim10^3$~km~s$^{-1}$.

\end{abstract}

\section{Introduction}

A number of high inclination (i.e. close to edge-on) low-mass X-ray binaries 
(LMXB) show regular and irregular flux decrease that occurs 
at every orbital cycle. 
The former is eclipse of a central compact source by a companion low-mass star 
and the latter, called as "dip" is 
thought to obscuration by thickened outer region of accretion disk where 
accretion stream from the companion star impacts \citep{white82}. 
Another feature responsible for the 
gas near a neutron star is narrow absorption line from highly ionized ions. 

The spectral changes during dip are complicated, and cannot be reproduced 
with simple absorption with neutral materials. 
Two component model consisting a compact blackbody and an extended 
power-law has been succeeded in explaining the dip and persistent spectra 
\citep{parmar86, church95, church97}. 
The compact source is heavily absorbed during the dip, while the 
extended component is gradually absorbed as increasing dip flux. 
Based on the two-component model, \citet{boirin05} and \citet{diaztrigo06} 
proposed an alternative interpretation that the spectra are simply 
explained with ionized absorber using the updated photo-ionization code. 

Narrow absorption lines have been also used to diagnose the gas 
property surrounding a compact object, 
such as the abundances, ionization states, the photo-ionization 
parameters or plasma temperatures of the gas.  
Accordingly, an X-ray source that exhibits the eclipse, dips and 
absorption lines is unique and powerful to study the high inclination 
LMXB systems. 
However only two LMXBs, MXB\,1659--298 and EXO\,0748--676 
(\cite{sidoli01}, \cite{parmar86}), 
has been known to show both the eclipse and dip. 
Among them, MXB\,1659--298 is unique showing narrow absorption lines 
from highly ionized atoms. The profiles of the narrow lines in MXB\,1659--298 
is consistent with resonance scattering by photo-ionized plasma, 
but shows no dependence on the dip and persistent flux.  

\begin{table*}[hbt]
  \begin{center}
    \label{tabl:obslog}
    \caption{Observation log.}
    \begin{tabular}{lccccc}
      \hline
      Start Date&Observatory&Obs. ID&\multicolumn{2}{c}{Aim point (J2000.0)}&$t_{\rm exp}$\footnotemark[$\dagger$]\\
      &&&R.A.&Decl.&(ks)\\
      \hline
      2005-09-23\footnotemark[$*$]& Suzaku&100027010  &\timeform{17h46m04s}&--\timeform{28D55'36''}&49\\
      2005-09-30\footnotemark[$*$]& Suzaku&100037040  &\timeform{17h46m04s}&--\timeform{28D55'36''}&47\\
      2007-07-19& Chandra&  8567  &\timeform{17h47m22s}&--\timeform{28D11'36''}&20\\
      2007-09-03& Suzaku&102013010  &\timeform{17h46m03s}&--\timeform{28D55'40''}&58\\
      \hline
      \multicolumn{4}{@{}l@{}}{\hbox to 0pt{\parbox{120mm}{\footnotesize
            \par\noindent
            \footnotemark[$*$] Observations used to construct the background. 
	    \par\noindent
            \footnotemark[$\dagger$] Exposure time after removing periods with
            high background level. 
	    Those of XIS are shown for Suzaku observations. 
          }\hss}}
    \end{tabular}
  \end{center}
\end{table*}

\axj\ was discovered in the ASCA observations of the Galactic center
\citep{maeda96, kennea96}. Showing a type-I X-ray burst and eclipses, 
it was classified as an eclipsing LMXB at a distance of $\sim 10$~kpc.
Except the X-ray burst and eclipses, the light curve was
nearly constant \citep{maeda98}.
The X-ray spectrum during the persistent state was featureless,
and reproduced by highly absorbed ($\nh\sim2.5\times10^{23}$~cm$^{-2}$)
power-law with a photon index of $\sim 2.4$, or thermal bremsstrahlung
with a temperature of $\sim 7.4$~keV.
The orbital period derived from the periodic eclipse
was $8.356\pm 0.008$~hr \citep{maeda96}. 

We discovered deep dips and narrow absorption lines from \axj\ with the 
Suzaku observations.  
Thus \axj\ is the second LMXB that exhibits the eclipse, dips and 
absorption lines. 
Furthermore, Suzaku provides us with the best-quality X-ray spectra in the 
three phases (persistent, dip, and 
eclipse).  This paper reports a unified spectral analysis of the three 
phases in the combination of the narrow line analysis, and then presents 
a new view of the high inclination LMXB.

\section{Observation and Data Reduction}

The Suzaku satellite \citep{mitsuda07} has carried out three observations 
of a field containing the Galactic nucleus Sgr~A$^*$ 
in 2005 September, 2006 September, and 2007 September. 
In the observation conducted on 2007 September 3--5, \axj\ 
was detected in outburst. 
The observation log is given in table~\ref{tabl:obslog}. 

\medskip

Suzaku has X-ray Imaging Spectrometer 
(XIS: \cite{koyama07a}), consisting four X-ray CCD cameras 
each placed on the focal planes of the 
X-Ray Telescope (XRT: \cite{serlemitsos07}). 
All four XRT modules are co-aligned to image the same region of 
18\arcmin$\times$18\arcmin\ with a half power diameter of 
1\farcm9--2\farcm3. 
Three of the cameras (XIS\,0, XIS\,2, and XIS\,3) have front-illuminated (FI) 
CCDs sensitive in the 0.4--12~keV energy band and the remaining one (XIS\,1) 
has back-illuminated (BI) CCD sensitive in the 0.2--12~keV energy band. 
XIS\,2 has been dysfunctional 
due to an anomaly occurred in 2006 November. 
Combined with XRT, the total effective area is $\sim400$~cm$^2$ at 8~keV. 
To mitigate the energy resolution degradation caused 
by charged particle radiation, 
the XIS has been equipped with Spaced-row Charge Injection (SCI). 
The details of this capability are described in 
\citet{bautz04, bautz07, uchiyama07}. 
The SCI technique was used in the observation of 2007 September. 
We checked the energy scale and resolution using the Mn~K$\alpha$ 
line from the calibration sources $\left(\atom{Fe}{}{55}\right)$ 
illuminating two corners of each CCD. 
As a result, we confirmed that the uncertainty of absolute energy scales are 
less than 5~eV and the energy resolutions in the full width at 
half maximum (FWHM) at 5.9~keV are $\sim145$~eV and $\sim180$~eV 
for FI and BI CCDs, respectively. 

Suzaku also has a non--imaging Hard X-ray Detector 
(HXD: \cite{kokubun07, takahashi07}). 
The HXD is comprised of Si PIN diodes (PIN) sensitive in 10--60~keV 
and GSO scintillator (GSO) sensitive in 40--600~keV, both located inside 
an active BGO shield. 
The PIN has a field of view (FOV) of 34\arcmin$\times$34\arcmin\ 
with lowest non--X-ray background ever achieved. 
We use only XIS and PIN data in this paper, because \axj\ is below 
the detection limit of GSO. 

The XIS and PIN data were processed with version 2.1. 
We removed data taken during the South Atlantic Anomaly passages 
and at earth elevation angles below 4\degree. 
For PIN data, we further excluded the events taken with 
cutoff rigidity less than 8~GV. 
We corrected for the dead time $(\sim5$--7\%) of the HXD 
using \texttt{hxddtcor}. 
The net exposure time after this filtering is $\sim 58$~ks for the XIS
and $\sim42$~ks for the PIN. 

\smallskip

We also retrieved archived Chandra \citep{weisskopf02} data to study 
temporal behavior of \axj. 
We found an observation using the Advanced CCD Imaging Spectrometer
(ACIS: \cite{garmire03}) covers a field containing \axj\ in 
outburst a month and half prior to 
the Suzaku observation (table~\ref{tabl:obslog}). 
For both Suzaku and Chandra data, barycentric corrections were 
applied. 

\section{Analysis and Results}
\subsection{Light curve}
\axj\ is only $\sim1\farcm5$ away from Sgr A$^*$. 
Since the bright supernova remnant Sgr~A~East \citep{maeda02} contaminates
to the source region and both the Galactic center diffuse X-ray emission 
and point source population are strongly peaked at the Galactic center 
\citep{koyama89, muno03, koyama07b}, 
the background photons cannot be accumulated from the nearby regions. 
Instead, we extracted them with the same region 
observed in 2005 September when the source was not in outburst. 
We simulated a point source image at the source position using 
\texttt{xissim} \citep{ishisaki07}, and extracted source photons at various 
enclosed photon fractions. 
Consequently, we found that a 48\% enclosed photon polygon maximize the 
signal-to-noise ratio. 
For the ACIS data, source photons were extracted from a 
20\arcsec$\times$40\arcsec\ elliptical region. 

\begin{figure*}[!hbt]
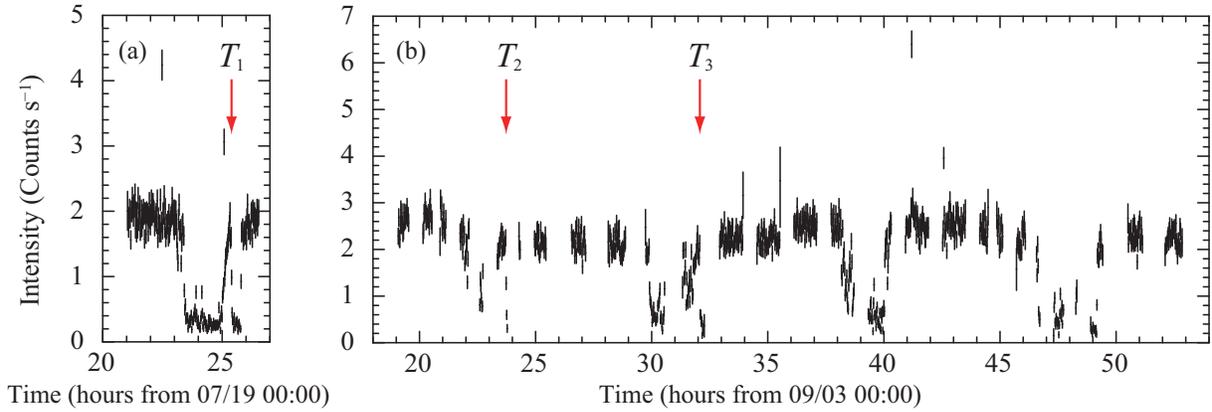

  \begin{center}
    \FigureFile(160mm,160mm){figure1.eps}
    \caption{Light curve of \axj\ in the 3.0--10.0~keV energy 
      band with a bin time of 80~s. The red arrows indicate the eclipse 
      ingresses. Data jumps are due to type-I X-ray bursts.
      (a) ACIS  light curve, where background is NOT subtracted. 
      (b) XIS light curve of combined data from the three XIS cameras, 
      where background is subtracted. }
    \label{fg:suzakulc1}
  \end{center}
\end{figure*}

\begin{figure*}[!htb]
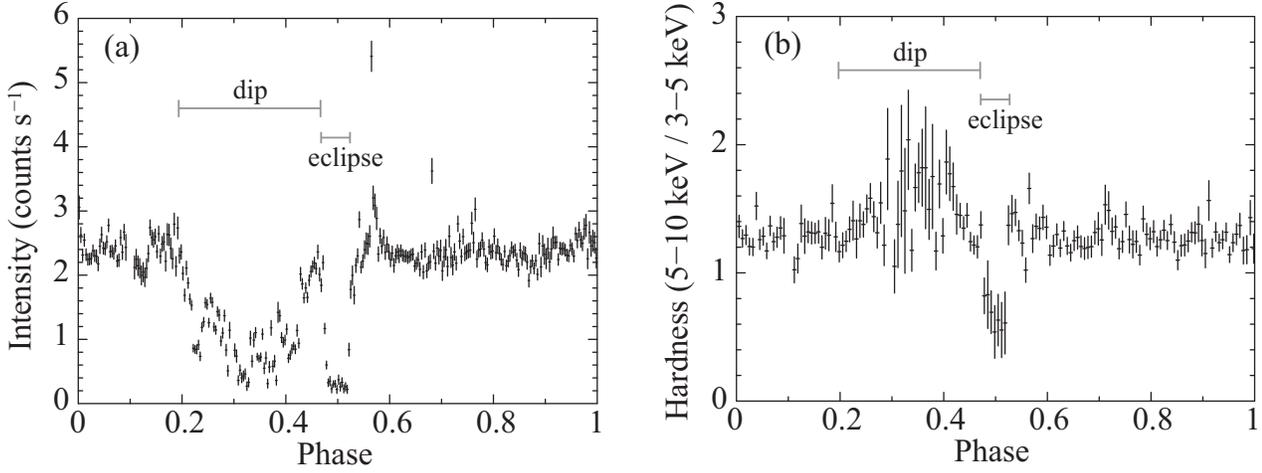

  \begin{center}
    \FigureFile(80mm,80mm){figure2a.eps}
    \hspace{0.5cm}
    \FigureFile(80mm,80mm){figure2b.eps}
  \end{center}
  \caption{(a) XIS 3-10~keV light curve folded with the orbital period of 
    30063.76~s.  Data jumps are due to type I X-ray bursts. (b) Hardness ratio 
    (counts in the 5--10~keV band divided by counts in the 3--5~keV band) 
    of the folded light curve. 
    In both panels, the background is subtracted.
  }\label{fg:foldsuzaku}
\end{figure*}

Figure~\ref{fg:suzakulc1} shows the ACIS and XIS light curves of \axj\ in 
the 3.0--10.0~keV energy band. 
The X-ray flux of \axj\ was highly variable; 
we found four type-I X-ray bursts, three fast flux drops, and two 
flux rises, which were already found by \citet{maeda96}. 
The presence of type-I bursts confirm that \axj\ is an LMXB.
The flux drops and rises are due to the eclipse of a companion star. 
We refer to the ingress times as $T_1$, $T_2$, and $T_3$ 
(figure~\ref{fg:suzakulc1}). 
Because the eclipse is expected to occur at every orbit 
with constant phase and duration, we assume
that $T_2-T_1=nP_{\rm orb}$\ and $T_3-T_2=P_{\rm orb}$, 
where $n$ is a natural number and $P_{\rm orb}$ is the orbital period. 
Under this condition, 
we obtained $P_{\rm orb}=30063.76\pm0.14$~s ($8.35104\pm 0.00004$~hour). 
Figure~\ref{fg:foldsuzaku} (a) shows the light curve folded with the 
orbital period 
and (b) shows the hardness ratio 
(counts in the 5--10~keV energy band divided by those in 
the 3--5~keV energy band). 
The phase origin point is so defined that the center of eclipse 
is 0.5. 
In addition to the above mentioned variability, we found relatively slow 
intensity dips in prior to the eclipses for the first time. 

We here define three states of \axj\ : ``persistent'' with orbital phase 
of 0--0.193 and 0.524--1, ``dip'' with orbital phase of 0.193--0.476,
``eclipse'' with orbital phase of 0.476--0.524 (figure~\ref{fg:foldsuzaku}a). 
The spectrum in the eclipse phase is softer, while that in the dip phase
is harder on average than that in the persistent phase. 

\subsection{Persistent and Eclipse spectra}

In the spectral analysis, we concentrate on the Suzaku data 
having a better energy resolution and statistics than those of 
Chandra. 
The ancillary response files (ARFs) and response matrix files (RMFs) 
were produced using \texttt{xissimarfgen} \citep{ishisaki07} 
and \texttt{xisrmfgen}. 
Since FI CCDs have almost the same ARFs and RMFs, 
the XIS\,0 and XIS\,3 data were summed and simultaneously 
fitted with the XIS\,1 spectrum. 
\begin{figure}[!htb]
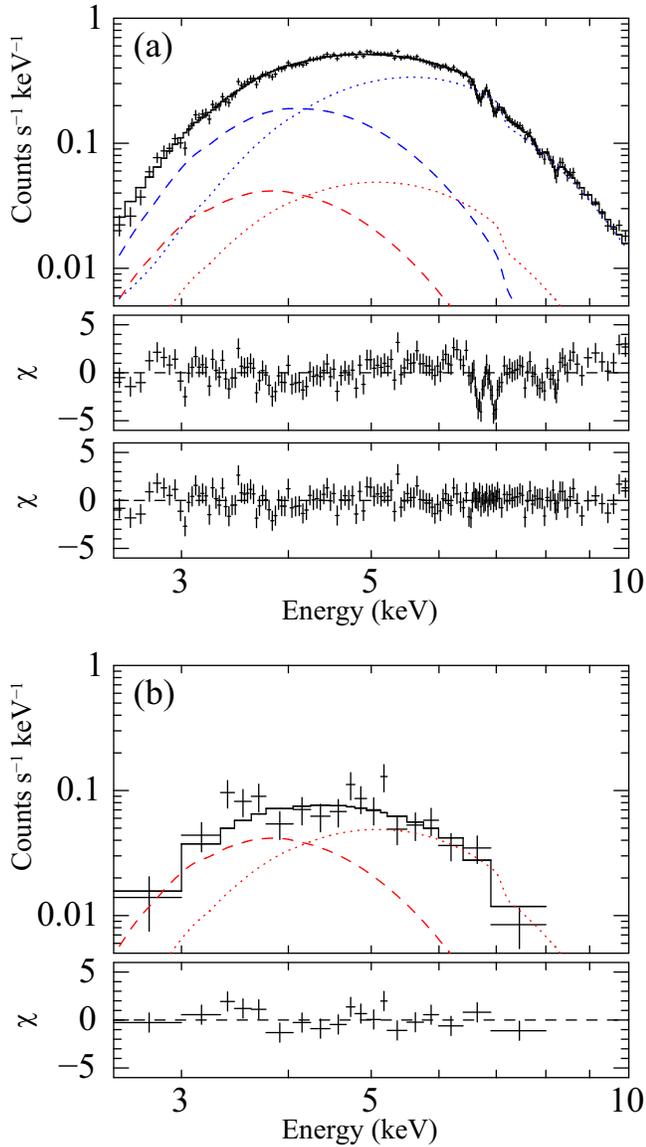

  \begin{center}
    \FigureFile(85mm,160mm){figure3a.eps}
    \FigureFile(85mm,160mm){figure3b.eps}    
  \end{center}
  \caption{(a) Background-subtracted spectrum of the persistent phase 
　　in the 2.5--10.0~keV band. Only the summed FI spectrum is shown for brevity. 
    The top panel shows the data in crosses, the best-fit model with negative
    Gaussians in solid lines, the disk-blackbody in dashed lines, 
　　the blackbody in dotted lines, the scattered component in red and 
    direct components in blue, respectively. 
    The middle and bottom panels show the residuals between data and 
    the best-fit model without and with Gaussians.
    (b) Same as (a), but of the eclipse state. 
    }
  \label{fg:eprsisspec}
\end{figure}

\citet{maeda96} reported that the X-ray spectrum during the eclipse 
is interpreted as the dust-scattering; a portion of X-rays emitted from 
the source in small off-set angle from the line of sight are scattered 
by the interstellar dust into the line of the source sight. 
We here call as the "scattered-in" component. 
On the other hand, vice-versa scattering process should exist; 
a portion of X-rays in the line of sight is "scattered-out" 
of the line of the source sight. 
We therefore fitted the persistent and eclipse spectra simultaneously 
as below. 

\begin{eqnarray*} 
\hspace{-0.4cm}
\begin{array}{ll}
\left[\alpha\cdot{\rm e}^{-\tau(E)}+\beta(E)\cdot\left\{1-{\rm e}^{-\tau(E)}\right\}\right]\cdot F(E)&\rm{:Persistent} \\
\beta(E)\cdot\left\{1-{\rm e}^{-\tau(E)}\right\}\cdot F(E)&\rm{:Eclipse} \\
\end{array} 
\end{eqnarray*}

, where $E$ is the photon energy in keV, $\alpha\equiv 0.48$ is the flux ratio
coming in the source region (section 3.1), and $\beta(E)$ is the fraction of
the extended (dust scattering) halo coming into the source region. Unlike $\alpha$, $\beta(E)$
depends on the X-ray energy ($E$).
$\tau(E)$ is the dust-scattering  optical depth, and $F(E)$ is the source spectrum. 
For $\tau(E)$, we assumed that the scattering optical depth is proportional 
to the inverse square of the photon energy, i.e., 
$\tau(E)=\tau_{\rm 1keV}\cdot E^{-2}$, 
where $\tau_{\rm 1keV}$ is the scattering optical depth at 1~keV. 
The first term of the persistent flux (upper equation) indicates that lower energy 
photons are scattered-out more than higher energy photons.  The second term is 
scattered-in photons. 

In order to estimate $\beta(E)$, we modeled the dust-scattered halo size
of \axj\ using another X-ray binary near the Galactic center \gs\ as follows: 
(1) \citet{mitsuda90} measured the energy-dependent size of 
the dust scattering halo of \gs\ using the lunar occultation. 
Assuming that the halo size is proportional to the inverse of the 
photon energy, we made the two-dimensional halo extensions 
with an energy step of 0.25~keV. 
(2) We convolved the halo images with response of the XRT+XIS 
using \texttt{xissim} at each energy step. 
(3) Extracting the photons from the source region in the simulated images, 
we obtained the energy-dependent enclosed photon fractions and fitted with a 
cubic function.  
(4) We implemented the explicit form as $\beta(E)= 
\left(-0.02259+0.03046\times E+0.00145\times E^2-0.00016\times E^3\right)$
to XSPEC. 

For $F(E)$, we used the standard model of low-mass binary: 
disk-blackbody (DBB) plus blackbody 
\citep{mitsuda84, makishima86} both attenuated by a common 
interstellar extinction \citep{morrison83} with solar abundances 
\citep{anders89}. 
This simple model however was rejected 
with $\chi^2$/d.o.f. (degree of freedom$)=556.0/335$ 
(figure~\ref{fg:eprsisspec}). 
The large $\chi^2$ value is mainly attributed to narrow negative 
residuals around 6.6--8.3~keV. 
Adding four negative Gaussian lines to the model, we obtained 
an acceptable fit with $\chi^2/{\rm d.o.f.}=337.4$/327. 
The widths of lines were consistent with zero, 
and the 95\% upper limit is 40~eV for 1~$\sigma$ in the 6.7~keV line. 
Allowing the iron abundance ($Z_{\rm Fe}$) in the interstellar absorber to 
be free, we obtained a better fit with $\chi^2/{\rm d.o.f.}=310.4$/326. 
The best-fit parameters are shown in table~\ref{tabl:xispersis}. 

\begin{table}[!ht]
  \begin{center}
    \caption{The best-fit parameters of the persistent and eclipse emission.}
    \label{tabl:xispersis}
    \begin{tabular}{llr@{}l}
      \hline
      Component&Parameter(Unit)&\multicolumn{2}{c}{Value\footnotemark[$*$]}\\
      \hline
      \multicolumn{4}{c}{Continuum}\\
      \hline
      Dust-scatter & $\tau_{\rm 1keV}$ &10.5&$^{+0.4}_{-0.8}$\\
      Absorption & $N_{\rm H}$~(10$^{23}$~cm$^{-2})$ & 1.87&$\pm 0.02$\\
      &$Z_{\rm Fe}$~(solar)&1.26&$^{+0.18}_{-0.16}$\\
      DBB& $kT_{\rm in}$\footnotemark[$\dagger$]~(keV)&0.72&$\pm 0.06$\\
      &${R_{\rm in}}^2\cos i$~(km$^2$)\footnotemark[$\ddagger$]&230&$\pm 5$\\
      Blackbody& $kT$~(keV)&1.64&$^{+0.05}_{-0.06}$\\
      &Area~(km$^2$)\footnotemark[$\ddagger$]&4.02&$\pm 0.04$\\
      \hline
      \multicolumn{4}{c}{Absorption lines}\\
      \hline
      Line 1&Line center~(eV)&6690&$^{+15}_{-17}$\\
      &EW~(eV)&$-49$&$^{+7}_{-6}$\\
      Line 2 &Line center~(eV)&6969&$^{+14}_{-11}$\\
      &EW~(eV)&$-57$&$\pm 7$\\
      Line 3 &Line center~(eV)&7866&$^{+57}_{-54}$\\
      &EW~(eV)&$-31$&$^{+13}_{-12}$\\
      Line 4&Line center~(eV)&8192&$^{+44}_{-45}$\\
      &EW~(eV)&$-36$&$^{+14}_{-15}$\\
      \hline
      \fx$_{\rm , persistent}$\footnotemark[$\S$]&(10$^{-11}$~erg~s$^{-1}$~cm$^{-2}$)&\multicolumn{2}{c}{12}\\
      \fx$_{\rm , eclipse}$\footnotemark[$\S$]&(10$^{-11}$~erg~s$^{-1}$~cm$^{-2}$)&\multicolumn{2}{c}{1.3}\\
      $\chi^2$/d.o.f.&&\multicolumn{2}{c}{310.4/326}\\
      \hline
      \multicolumn{4}{@{}l@{}}{\hbox to 0pt{\parbox{100mm}{\footnotesize
	\par\noindent
	\footnotemark[$*$] The errors are at 90\% confidence level.\\
	\footnotemark[$\dagger$] Temperature at the innermost radius.\\
	\footnotemark[$\ddagger$] A distance of 10~kpc is assumed.\\
	\footnotemark[$\S$] Observed energy flux in the 3.0--10.0~keV band.\\
	\footnotemark[$\|$] Extinction-corrected luminosity in the 3.0--10.0~keV band.\\
	  }\hss}}
    \end{tabular}
  \end{center}
\end{table}

The best-fit line center energies, $\sim6.69$~keV, $\sim 6.97$~keV, 
$\sim7.87$~keV, and $\sim8.19$~keV indicate that these are due to 
resonance scattering by highly ionized iron and/or nickel 
as seen in other LMXBs 
\citep{ueda01, sidoli01, parmar02, boirin03, boirin04, boirin05, church05}. 

We confirmed that any line is significant based on $F$-test. 
For example, even $\sim8.19$~keV, having the worst statistics among 
the four lines, is significant at more than 99.99\% confidence level 
with an $F$-value of 9.66. 
Fe\emissiontype{XXV} and Fe\emissiontype{XXVI} absorption edges at 
8.828~keV and 9.278~keV should be accompanied with 
the highly ionized iron absorption lines. 
Then, adding \texttt{edge} model of XSPEC, we obtained an upper limit 
of $\sim0.1$ for the optical depths at around 9~keV. 

\subsection{Dip spectra}
\begin{figure}[!htb]
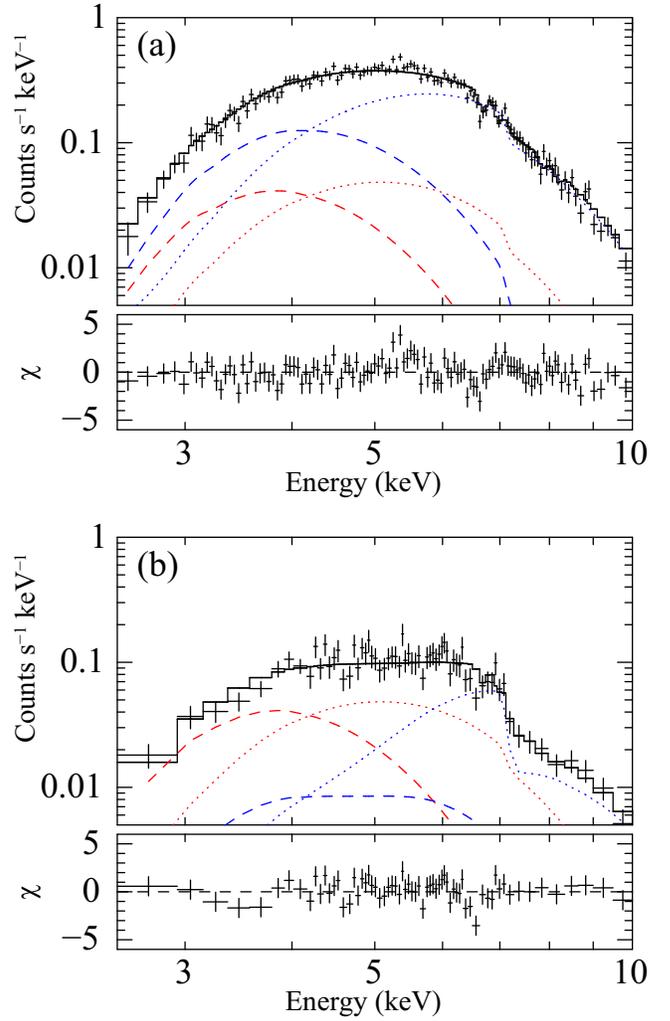

  \begin{center}
    \FigureFile(85mm,160mm){figure4a.eps}
    \FigureFile(85mm,160mm){figure4b.eps}
  \end{center}
  \caption{(a) Same as figure~\ref{fg:eprsisspec}, 
    but for the shallow dip state. 
    (b) Same as figure~\ref{fg:eprsisspec}, 
    but for the deep dip state. 
  }
  \label{fg:shdipspec}
\end{figure}

To investigate the spectral shape and its variation during the dip, 
we divided the dip into two phases:
``shallow dip'' with the XIS count rates more than 0.7~counts~s$^{-1}$ 
and ``deep dip'' with those less than 0.7~counts~s$^{-1}$. 
We first simply assume that the dip spectra are comprised of 
the scattered and direct components, and the latter is partially absorbed 
by cold matter. 
We fitted the spectra with the form,\\
 
$\left\{\alpha\cdot PC(E)\cdot {\rm e}^{-\tau(E)}+\beta(E)\cdot(1-{\rm e}^{-\tau(E)})\right\}\cdot F(E)$. \\
Here $PC(E)$ is a partial covering model given as,\\ 

$PC(E)=f\cdot{\rm e}^{-\nh\cdot\sigma(E)}+\left(1-f\right)$,\\ 
where $f$ is the covering fraction ($0\le f\le1$), and $\sigma(E)$ is 
the cross section of the absorbing material. 
All the parameters included in $F(E)$ were fixed to those 
of the best-fit values in  the eclipse and persistent phase (see section 3.2). 

\begin{table}[!htb]
  \begin{center}
    \caption{The best-fit parameters of the dip state
      with a cold partial covering absorber model.}
    \label{tabl:dippar}
    \begin{tabular}{lcc}
      \hline
      Parameter (Unit) &\multicolumn{2}{c}{\hrulefill Value\footnotemark[$*$]\hrulefill}\\
      &Shallow dip&Deep dip\\
      \hline
      $f$&0.35$\pm 0.03$&0.96$\pm 0.01$\\
      \nh\ (10$^{23}$~cm$^{-2}$)&6.1$^{+1.8}_{-1.2}$&10.4$^{+0.9}_{-0.8}$\\
      \fx\ (10$^{-11}$~erg~s$^{-1}$~cm$^{-2}$)&9.1&2.8\\
      $\chi^2$/d.o.f.&264.9/225&95.6/89\\
      \hline
      \multicolumn{3}{@{}l@{}}{\hbox to 0pt{\parbox{85mm}{\footnotesize
            \par\noindent
            \footnotemark[$*$]    The errors are at 90\% confidence level.\\
          }\hss}}
    \end{tabular}
  \end{center}
\end{table}

We further examined the upper limit of ionization state of the partial covering absorber. 
We used \texttt{absori} model \citep{done92,magdziarz95} in XSPEC and 
found that the 90\% upper limit of ionization parameter 
$\xi(=\frac{L}{nr^2})$ is 4.7 
and 0.03 for the shallow and deep dip states, respectively. 
Alternatively, we fixed the iron abundance in the absorber to 0 and 
added absorption edge. Then we obtained the edge energy, 
7.25$^{+0.19}_{-0.12}$~keV and 7.16$^{+0.08}_{-0.07}$~keV 
in the shallow and deep dip state respectively. 

\subsection{XIS+HXD analysis}

\begin{table}[!ht]
  \caption{The best-fit parameters for the "difference spectrum" 
(persistent minus eclipse spectra) of the combined spectra of XIS and HXD.} 
  \begin{center}
  \begin{tabular}{lcr@{}l}
    \hline
    Component & Parameter (Unit) & \multicolumn{2}{c}{Value\footnotemark[$*$]}\\
    \hline
    Absorption & $N_{\rm H}$ (10$^{23}$~cm$^{-2}$) &1.72&$^{+0.06}_{-0.05}$ \\
    & $Z_{\rm Fe}$ (solar) & 1.32&$\pm 0.18$\\
    DBB & $kT_{\rm in}$~(keV)&0.70&$\pm0.07$\\
    &${R_{\rm in}}^2\cos i$~(km$^2)$&190&$\pm20$\\
    Blackbody& $kT$ (keV)&1.64&$\pm0.07$\\
    &Area (km$^2$)&4.14&$\pm0.04$\\
    \fx&(10$^{-11}$~erg~s$^{-1}$~cm$^{-2}$)&\multicolumn{2}{c}{11}\\
      $\chi^2$/d.o.f.&&\multicolumn{2}{c}{150.8/150}\\
    \hline
    \multicolumn{4}{@{}l@{}}{\hbox to 0pt{\parbox{75mm}{\footnotesize
	  \par\noindent
	  \footnotemark[$*$]    The errors are at 90\% confidence level.\\
	  \footnotemark[$\dagger$] Absorption-corrected luminosity in the $3-10$~keV band. 
	  A distance of 10~kpc is assumed.\\
	}\hss}}     
  \end{tabular}
  \label{tabl:xishxd}
  \end{center}
\end{table}

\begin{figure}[!htb]
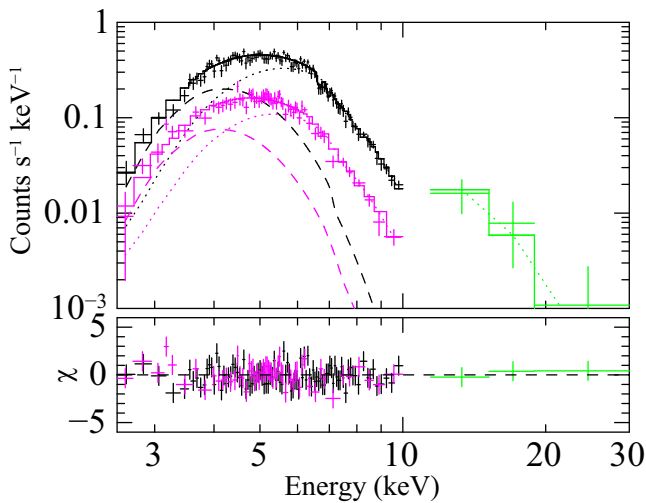

  \begin{center}
    \FigureFile(85mm,160mm){figure5.eps}
  \end{center}
  \caption{Same as figure~\ref{fg:eprsisspec}, 
    but the background is extracted from the eclipse state.
    Top panel shows the summed FI data in black, the BI data in magenta, 
    and the HXD data in green. 
    }
  \label{fg:xishxdspec}
\end{figure}

\axj\ is in the high background area of the Galactic centre diffuse X-ray 
(GCDX) \citep{koyama96, koyama07b, yuasa08}. 
Also transient or time variable sources may be in the PIN FOV.  
To avoid systematic errors due to the GCDX and transient/time variable 
sources, we made a "difference-spectrum" of PIN. 
We first excluded the non-X-ray background (NXB) signals of the PIN 
from the persistent and eclipse state respectively, 
and subtracted the NXB-excluded eclipse spectrum from the NXB-excluded 
persistent spectrum. 

Since \axj\ resides apart from the optical axis of the HXD by 
$\sim7$\arcmin\ in the present observation attitude, 
the effective area of PIN decreases by $\sim20$\%. 
We calculated the corresponding ARF using \texttt{hxdarfgen}, and 
applied it in the spectral analysis. 
Making the XIS spectrum in the same way, we performed the simultaneous 
fitting to the XIS and HXD data. 
We fixed the PIN/XIS cross-normalization factor to 1.15, the mean value of 
luminous sources 
\citep{kokubun07, reeves07, miniutti07, shirai08, markowitz08}. 

The scattered-in component should be automatically subtracted in this method, 
and hence only the direct component is taken into account. 
Since the photon statistics is limited, the absorption line parameters 
and $\tau_{rm 1keV}$, responsible for scattering-out, are fixed to those 
derived in \S 3.2. 
The XIS and HXD spectra are well reproduced by almost the same parameters 
in the direct component in \S 3.2 (figure~\ref{fg:xishxdspec}). 
As is seen in figure\ref{fg:xishxdspec}, the hard band data 
($\sim10$--30~keV) is simply described by the blackbody spectrum of 
$kT=1.64\pm0.07$~keV, which is most likely
due to the emission of a neutron star surface. No hard tail (power-law) 
component such as the Compton-up scattering emission is requited. 

\section{Discussion}
\subsection{Persistent Spectra}

\axj\ is the second low-mass X-ray binary that exhibits dips, eclipse and 
X-ray burst, and absorption lines. 
Using the best-quality Suzaku spectra of \axj, we 
found a clearer picture of this LMXB type.
The X-ray spectrum in the persistent phases are described with a 
combination of direct, 
and dust scattering (scattered-in and scattered-out) emissions. 
The direct component is composed of DBB 
and a blackbody (neutron star surface).
Although dipping/eclipsing sources are  high inclination (nearly disk-edge on) 
systems, we need no power-law components in the hard energy band up to 30 keV, 
and hence Compton-up scattering 
by possible hot disk corona is absent (see section 4.3).

\subsection{Eclipse Spectra}

The eclipse spectrum is purely dust scattering (scattered-in) component with
a large optical depth ($\tau_{\rm 1keV}$) of 10.
\citet{predehl95} conducted a systematic study on interstellar absorption 
and dust scattering using the ROSAT data. 
They found $\tau_{\rm 1keV}$ and \nh\ correlates well with 
$\nh/\tau_{\rm 1keV}\sim2\times10^{22}$~cm$^{-2}$ in the $\nh$ range of 
0--3$\times10^{22}$~cm$^{-2}$. Our data for \axj\ 
$\nh/\tau_{\rm 1keV}\sim 1.8\times10^{22}$~cm$^{-2}$ 
extends this relation over nearly one order of magnitude of $\nh$,
up to $\sim2\times10^{23}$~cm$^{-2}$. 
Conversely, this good correlation strongly supports our assumption that 
the eclipse emission 
is due to dust scattering. Also the gas-to-dust ratio towards the 
Galactic center is not 
largely different from the other regions in the Galaxy. 

\subsection{Dip Spectra}

The spectra are simply explained by a partial covering model in both the 
deep and shallow dips. 
No extended emission of power-law spectrum such as Compton-up scattering 
is required.
The flux decrease is mainly due to the increase of the covering factor, 
together with a slight increase
of \nh. Unlike the prediction of \citet{diaztrigo06}, the absorption gas is 
not highly ionized. 
The upper limit of photo-ionization parameter $\xi(=\frac{L}{nr^2})$ is 
0.03--4.7, depending on the depth of dip.

\subsection{Absorption lines}

We found four absorption lines with the respective line center energies and 
equivalent widths of (1)  $6690^{+15}_{-17}$~eV and $49^{+7}_{-6}$~eV, 
(2) $6969^{+14}_{-11}$~eV and $57\pm 7$~eV, (3) $7866^{+57}_{-54}$~eV and 
$31^{+13}_{-12}$~eV, and (4) $8192^{+44}_{-45}$~eV and $36^{+14}_{-15}$~eV.
Since the absorption line features are found in every orbital phases, these
are due to disk corona gas. 
From the center energies of the absorption lines, we infer that (1) and 
(2) are Fe\emissiontype{XXV}~K$\alpha$ and 
Fe\emissiontype{XXVI}~K$\alpha$, respectively.
Line (3) would be a complex of Fe\emissiontype{XXV}~K$\beta$ and 
Ni\emissiontype{XXVII}~K$\alpha$, 
while line (4) is Fe\emissiontype{XXVI}~K$\beta$ plus 
Ni\emissiontype{XXVIII}~K$\alpha$.
The EW of resonance scattering absorption line is a function of 
both the column density 
and velocity dispersion along the line of sight ($\Delta v_{\rm los}$),
whereas the absorption edge depth of corresponding ionic species 
depends only on the column density. 
The upper limit of absorption edge ($\tau\leq 0.1$) gives upper limit 
of $\sim5\times10^{18}$~cm$^{-2}$ for $N_{\rm FeXXV}$. 
Combining the EW of the 6.7~keV line ($49$~eV), we constrain the lower 
limit of $\Delta v_{\rm los}\geq4\times10^2$ km~s$^{-1}$(see figure 3 in 
\cite{kotani06}).
On the other hand, the upper limit of 
$\Delta v_{\rm los}<2\times10^3$~km~s$^{-1}$ is obtained from the upper limit 
of the line width of 40 eV. 

For simplicity, we assume an intermediate value of 
$\Delta v_{\rm los}=7\times10^2$~km~s$^{-1}$ 
(kinematics temperature of $10^2$~keV, figure~3 by \cite{kotani06}), 
and discuss physical condition of iron and nickel by referring 
figure~3 of \citet{kotani06}. 
Using the EWs of Fe\emissiontype{XXV} and Fe\emissiontype{XXVI} lines of 
$49$~eV and $57$~eV, 
we estimate the column density of Fe\emissiontype{XXV} ($N_{\rm FeXXV}$) and 
Fe\emissiontype{XXVI} ($N_{\rm FeXXVI}$) to be $1.8^{+0.7}_{-0.6}\times10^{18}$~cm$^{-2}$ 
and $4.2^{+3.6}_{-2.5}\times10^{18}$~cm$^{-2}$, respectively. 
Then EWs of K$\beta$ of Fe\emissiontype{XXV} and Fe\emissiontype{XXVI} 
are estimated 
to be $20\pm5$~eV and $17^{+9}_{-6}$~eV respectively, and hence EWs of K$\alpha$ of 
Ni\emissiontype{XXVII} 
and Ni\emissiontype{XXVIII} are $<24$~eV and $19^{+10}_{-15}$~eV. 
The latter two values constrain 
$N_{\rm Ni\emissiontype{XXVII}}$ and $N_{\rm Ni\emissiontype{XXVIII}}$ to be 
$<4\times10^{17}$~cm$^{-2}$ and $4^{+4}_{-3}\times10^{17}$~cm$^{-2}$, respectively. 
The ionic fractions of Fe\emissiontype{XXVI} and Ni\emissiontype{XXVIII} 
are $0.68\pm0.08$ and $0.40\pm0.05$, so the abundance ratio of Ni/Fe is 
0.13$^{+0.13}_{-0.10}$, which is consistent with the solar value
\citep{anders89}. 
Assuming the solar abundance of iron, the $N_{\rm H}$ value for the 
absorption gas is $\sim1.3\times10^{23}$, or optical depth of Thomson 
scattering ($\tau_{\rm es}$) is $\sim0.1$. 
 
The ratio of $N_{\rm Fe\emissiontype{XXVI}}$/$N_{\rm Fe\emissiontype{XXV}}$ 
gives the plasma temperature of $13\pm4$~keV in collisional equilibrium
\citep{arnaud92}.
On the other hand, if the absorption gas is due to photo-ionized plasma, 
then the ionization parameter ($\xi$) is $\sim10^{3.5}$ and 
the plasma temperature is $\sim100$~eV.
Using the plasma temperatures given above, the random (thermal) velocity 
of iron/nickel estimated to be less than 
300 km~s$^{-1}$, significantly lower than 700 km~s$^{-1}$. 
Therefore the velocity of 700 km~s$^{-1}$ would be mainly attributable 
to a bulk motion of the gas. 

This velocity is very small as an electron velocity, therefore in any case, 
the electron temperature is less than 20~keV, and hence 
the $y$ parameter ($4\tau_{\rm es}\cdot kT$/511~keV) is less than 0.02.
This small value of $y$ 
is consistent with no Compton-up scattering 
flux in the hard energy band (section 4.1).  

\section{Summary}

We observed the eclipsing LMXB \axj\ in an outburst phase
with Suzaku. The results are summarized as follows:

\begin{enumerate}
\item  Combining the Chandra data obtained 1.5 month earlier than Suzaku, 
the orbital period is 
more accurately constrained than before to be $30063.76\pm0.14$~s.

\item Irregular dipping activity was found for the first time. 

\item Narrow absorption lines were detected at $\sim6.7$~keV, $\sim6.9$~keV, 
$\sim7.8$~keV, and $\sim8.2$~keV. 

\item The narrow line profiles are described by absorptions of K-shell 
transition of highly ionized iron and nickel in solar abundance. 

\item The persistent spectrum is well modeled with a direct and scattered in by 
dust component. 

\item The eclipse spectrum is comprised only of the scattered-in component.

\item The dip spectra are well reproduced by direct emission absorbed by cold 
matter and the scattered-in component. 

\item The HXD signal in the 12--30~keV band was detected by subtracting 
the eclipse spectrum from persistent spectrum. 
The emission was reproduced by a blackbody component with a temperature 
of $\sim1.6$~keV. No significant hard tail was detected.

\end{enumerate}
\bigskip

We thank Makoto Sawada and Taro Kotani for useful comments and discussions. 
Y.\,H. and T.\,Y. are financially supported 
by the Japan Society for the Promotion of Science. 
The work is financially supported by the grants-in-aid for a 
21st century center of excellence program 
``Center for Diversity and Universality in Physics'' and No. 18204015 
by the Ministry of Education, Culture, Sports, Science and Technology 
of Japan. 
This research has made use of data obtained from the Data ARchive and
Transmission System at ISAS/JAXA.

\end{document}